\documentclass[preprint,12pt]{elsarticle}
\usepackage{graphicx} 
\usepackage{amssymb}
\usepackage{amsmath}
\usepackage{mathtools}
\usepackage{bm}
\usepackage[hyphens]{url}
\usepackage{hyperref}
\usepackage{algorithm}
\usepackage{algpseudocode}
\usepackage{lineno}
\usepackage{booktabs}
\usepackage{adjustbox}
\usepackage{subcaption}
\usepackage{lipsum}
\usepackage{xcolor}
\usepackage{doi} 
\usepackage[utf8]{inputenc} 
\usepackage[T1]{fontenc}    
\usepackage{hyperref}       
\usepackage{url}            
\usepackage{booktabs}       
\usepackage{amsfonts}       
\usepackage{nicefrac}       
\usepackage{microtype}      
\usepackage{expl3} 

\journal{Reliability Engineering \& System Safety}

\begin{document}
\begin{frontmatter}


\title{Inverse uncertainty quantification of a mechanical model of arterial tissue with surrogate modelling}

\author[uvaaddress,utretchaddress]{Salome Kakhaia\corref{mycorrespondingauthor}}
\cortext[mycorrespondingauthor]{Corresponding authors, tel.: +31 647028787, email address: salome.kakhaia@gmail.com, postal address: Hijmans van Den Berghlaan 114, 3571 PD, Utrecht }

\author[uvaaddress,itmoaddress]{Pavel Zun}
\ead{Pavel.Zun@gmail.com}

\author[uvaaddress]{Dongwei Ye}
\ead{d.ye-1@utwente.nl}

\author[uvaaddress]{Valeria Krzhizhanovskaya}
\ead{v.krzhizhanovskaya@uva.nl}

\address[uvaaddress]{Computational Science Lab, Informatics Institute, Faculty of Science, University of Amsterdam, The Netherlands}

\address[itmoaddress]{National Center for Cognitive Research, ITMO University, Saint Petersburg, Russia}

\address[utretchaddress]{Julius Center
for Health Sciences and Primary Care, The University Medical Center Utrecht, The Netherlands}

\begin{abstract}
Disorders of coronary arteries lead to severe health problems such as atherosclerosis, angina, heart attack and even death. Considering the clinical significance of coronary arteries, an efficient computational model is a vital step towards tissue engineering, enhancing the research of coronary diseases and developing medical treatment and interventional tools. In this work, we applied inverse uncertainty quantification to a microscale agent-based arterial tissue model, a component of a multiscale in-stent restenosis model. Inverse uncertainty quantification was performed to calibrate the arterial tissue model to achieve the mechanical response in line with tissue experimental data. Bayesian calibration with bias term correction was applied to reduce the uncertainty of unknown polynomial coefficients of the attractive force function and achieved agreement with the mechanical behaviour of arterial tissue based on the uniaxial strain tests. Due to the high computational costs of the model, a surrogate model based on Gaussian process was developed to ensure the feasibility of the computation. 

\end{abstract}


\end{frontmatter}

\begin{table}[H]
\centering
\caption{Nomenclature}
\label{Tab1}
\resizebox{\columnwidth}{!}{
\begin{tabular}{|l|l|} 
\hline
$\textbf{z} $~                          	& Analytical stress response of uniaxial stress-strain tests   \\
$F_{bond}$~                                                 	& Attraction bond force                                        \\
$\varrho$~                                                  	& Bond strain                                                   \\
$\varepsilon_c$~                                            	& Code uncertainty in the IUQ model with variance $\sigma_c^2$                  \\
$\theta_i$~                                                 	& Coefficient of the attraction bond force                     \\
$d$~                                                        	& Distance between the centers of two interacting cells        \\
$\lambda$                                                   	& Extension ratio in the loading direction                     \\
$\epsilon$~                                                 	& Gaussian noise term of the surrogate model                   \\
$f_{GP}$~                                                   	& Gaussian process regression surrogate model                  \\
$\boldsymbol{x} = (\theta_1,...\theta_6,\varrho)$           	& Input of a surrogate model                           \\
$C_{ij}$~                                                   	& Material model parameter                                          \\
$\mu$ and $K$~                                              	& Mean and covariance functions of the surrogate model           \\
$\varepsilon_m$~                                            	& Model inadequacy in the IUQ model with variance $\sigma_m^2$                  \\
$S_{UT}$~                                                   	& Nominal (engineering) stress                                 \\
$N_{obs}$~                                                  	& Number of observed stress-strain values                      \\
$N_{tot}$~                                                  	& Number of total stress-strain values for testing interpolation capabilities              \\
$N_p$~                                                      	& Number of total surrogate model predictions~                              \\
$s = f_{GP}(\boldsymbol{x})$                                 	& Output of a surrogate model                      \\
$\Theta$~                                                   	& Parameter space of attractive force coefficients        \\
$\lambda_i$                                                 	& Principal extension ratio                                    \\
$\vartheta$                                                 	& Principal strain value                                       \\
$R1, R2$~                                                   	& Radii of two interacting cells                               \\
$\boldsymbol{\theta^*}$~                                    	& Reference value of the set of true attraction bond force coefficients  \\
$\varepsilon_r$~                                            	& Residual variability in the IUQ model with variance $\sigma_r^2$              \\
$f_{ABM}$~                                                  	& Response function of agent-based model of arterial tissue             \\
$\boldsymbol{\varrho}$                                       	& Set of bond strain values applied during uniaxial strain test     \\
$\boldsymbol{\theta}$                                       	& Set of coefficients of the attraction bond force             \\
$\Phi $                                                     	& Set of uncertain calibration parameters                 \\
$N_s$~                                                      	& Size of the sampled coefficient space                            \\
$N_t$~                                                      	& Size of the training data of the surrogate model             \\
$N_{val}$~                                                  	& Size of the validation data of the surrogate model               \\
$W $~                                                       	& Strain energy function                                       \\
$I_i$~                                                      	& Strain invariant                                             \\
$D = (\textbf{z},\textbf{s})$~                              	& Stress data for the IUQ model                                       \\
$\varepsilon$~                                              	& Total uncertainty in the IUQ model                               \\
\hline
\end{tabular}          
}
\end{table}

\pagebreak

\section{Introduction}
Cardiovascular diseases (CVDs) are the top cause of death globally, leading to an estimated 16\% of total deaths annually \cite{WHO}. Atherosclerosis, a narrowing or blocking of arteries due to a plaque buildup is a common CVD, which is often treated by performing percutaneous coronary interventions (PCIs) with stenting in coronary vessels {\cite{GALLO202294, Gaudino2023}}. However, an incidence rate of the excessive amount of neointima formation after PCI and a repeat narrowing of the vessel, a condition known as in-stent restenosis (ISR), is notably high, making it a public health concern of significant importance {\cite{Rempakos2023, Giustino2022}}. A three-dimensional computational model ISR3D has been developed to simulate smooth muscle cells proliferation and the ISR process, as well as to predict the restenosis progression under different scenarios \cite{Zun2017, Zun2019,ye2022}. 

ISR3D is a coupled multiscale model consisting of three single-scale submodels: the initial condition model, the vessel tissue model, and the blood flow model, as well as utility modules which facilitate communication between submodels \cite{Zun2017}. The vessel tissue submodel is a cell-scale agent-based model (ABM), and it uses pairwise repulsive and attractive forces between cell-representing agents to model the mechanical behaviour. Three layers of arterial tissue - diseased intima with atherosclerosis, media and adventitia are modelled explicitly. The microscale interactions result in a mechanical response representative of biological tissue on a macro scale. The attractive force incorporates the macroscopic properties into the cell-scale model. It is modelled as a $6^{th}$ order polynomial pairwise function, depending on the cell's size and the bond strain. The polynomial coefficients are model-specific, and no prior values of these coefficients are known. The goal of this work is to calibrate those unknown polynomial coefficients to match the macroscopic behaviour (i.e. the stress-strain experimental data of human coronary arteries) of the arterial tissue according to  \cite{Zun1, Holzapfel2005}.
The considered macroscopic data is the stress-strain relation for uniaxial strain tests of tissue stretched in a circumferential direction, which is a simple but effective representation of the strain occurring in the tissue during stenting or pressurisation. The existing continuum models with finite element (FE) methods have managed to capture the mechanical response of arterial tissues well \cite{Holzapfel2002, He2020}. Therefore, we utilized a continuum model to generate the data and subsequently applied them to ABM calibration. {By integrating continuum modelling, which provides a macroscopic view of the system with agent-based modelling, that allows for a more detailed understanding of the microscopic behaviour of individual agents on the cellular level \cite{Zun2019, Corti2020, Corti2022}, researchers can gain a better understanding of complex systems at multiple scales \cite{Corti2021}.}

{
Inverse uncertainty quantification (IUQ) is a process that determines the uncertain inputs based on the experimental data or measured output using Bayesian techniques \cite{Wu2021, Allen2022}. It allows people to integrate prior information of the uncertain inputs and update that knowledge with existing data based on the Bayes' rule \cite{von2011, Fox2012}. Since the marginal likelihood, used for normalizing the posterior distribution, is typically computationally intractable, multiple approximation methods have been developed to estimate the posterior distributions or the samples of the posterior distribution, such as Markov chain Monte Carlo (MCMC) \cite{Neal2011, Yang2022} and variation inference \cite{David2017}. The derived posterior distribution can be subsequently applied for predictive distribution. As a result, IUQ has found wide application in the field of computational science and engineering, particularly for solving inverse problems \cite{Saikumar2017, Majdi2019,MERLE2015305, Domitr2023}.}

{
IUQ typically requires a large number of model evaluations. Such many-query scenarios may become computationally prohibitive for a computationally expensive model. The problem can be addressed by employing surrogate modelling, that approximates the original complex model. One of the main categories of surrogates is data-driven methods, which consider the model as a black box and empirically approximate the latent mapping between inputs and model responses. Typical instances of data-driven methods include radial basis functions method \cite{Gutmann2001}, polynomial chaos expansion \cite{Lim2021, Zhou2022, He2022} , neural networks \cite{Bao2021, SEO2022108102}, support vector machines \cite{Roy2023}, etc. The Gaussian process is another state-of-the-art regression method and is widely applied due to its non-parametric and probabilistic nature \cite{ye2021, Saida2023, Wang2022}. It assumes that the responses of the model follow multivariate normal distributions and, conditioning on existing data, the response at a new input position can be predicted. The hyperparameters can be learned through maximum marginal likelihood. We developed the Gaussian process regression surrogate to substitute the original ABM for the evaluations in the inverse uncertainty quantification process.}

In this paper, we developed a Bayesian framework combining inverse uncertainty quantification techniques and a Gaussian process surrogate model to calibrate a microscale agent-based model of arterial tissue based on a macroscale continuum constitutive model for the mechanical behaviour of arterial tissue. This approach improved the model's accuracy, which serves as a submodel in the multiscale simulation of a complex process of in-stent restenosis. The reliability and accuracy of each component have the utmost importance for the overall reliability and precision of the complex system driven by the integration and interaction of the component modules.

The paper is arranged as follows. The material, agent-based and surrogate models of arterial tissue are described in Section~\ref{cha:methods}. In the same section, the IUQ process is as well introduced. The result of surrogate modelling and IUQ are presented in Section~\ref{cha:results}, followed by the discussion and conclusion in Section~\ref{cha:discussion} and \ref{cha:conclusion}. The nomenclature used in the paper is given in Table \ref{Tab1}.

\section{Methods}\label{cha:methods}
\subsection{Agent-based model of arterial tissue}
\begin{figure}[tb!]
\centering
 \begin{subfigure}{.45\textwidth}
  \includegraphics[width=\linewidth]{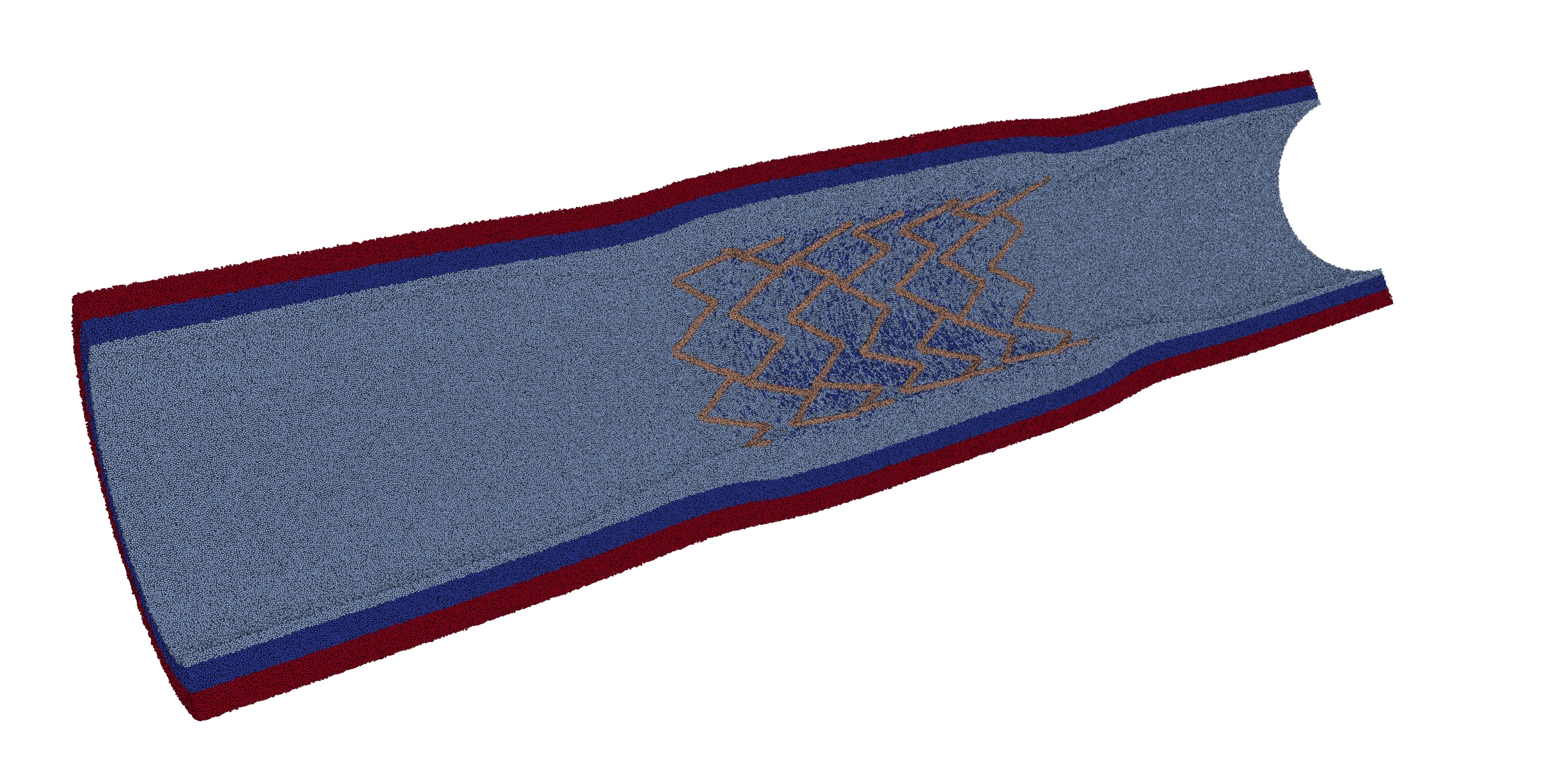}
  \caption{ABM of arterial vessel wall}
  \label{fig:vessel_wall}
\end{subfigure}
 \begin{subfigure}{.45\textwidth}
  \includegraphics[width=\linewidth]{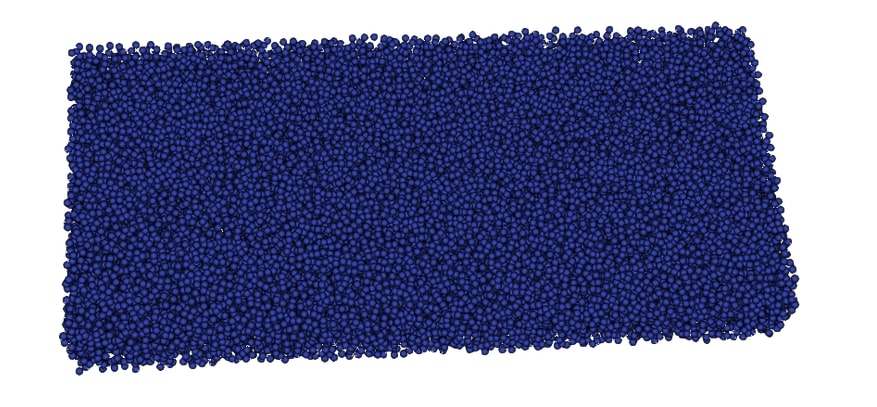}  
  \caption{Initial tissue layer}
  \label{fig:before_linear_force}
\end{subfigure}
\begin{subfigure}{.45\textwidth}
  \includegraphics[width=\linewidth]{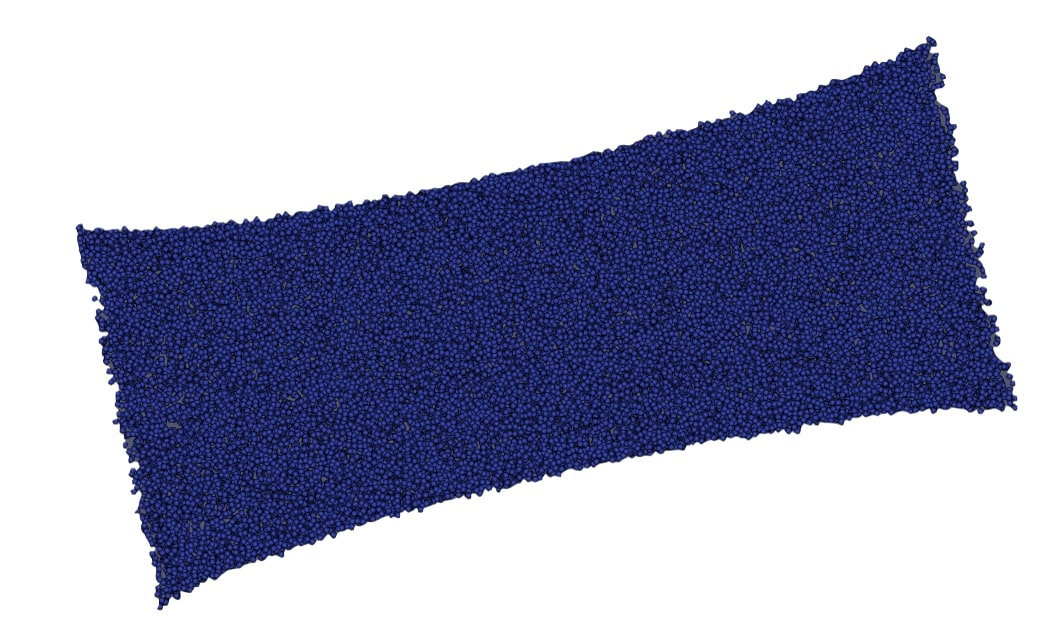}  
  \caption{Deformed tissue layer }
  \label{fig:after_linear_force}
\end{subfigure}
\caption{ (a) An illustration of the agent-based model of arterial vessel wall consisting of three layers: diseased intima (light blue), media (dark blue),  adventitia (red) and a metal stent. The single layer of arterial tissue before (b) and after (c) applying a particular value of uniaxial strain with the force coefficients $\theta_1=2.14$ and $\theta_2,...,\theta_6 = 0$. }
\label{fig:vessel_wall_deformation}
\end{figure}

The arterial tissue model is a part of the 3D multiscale model of in-stent restenosis (ISR3D), extensively described in \cite{Zun2019}.
From the mechanical point of view, the vessel wall is modelled as a centre-based agent-based model, where individual cells are modelled as spheres, interacting via adhesive and repulsive forces. Spheres have their radii set to match the volume of the cells, and forces act between cell centres. In ISR3D, biological behaviour is modelled by imposing a biological ruleset on each agent, while simplified mechanical properties of the centre-based approach allow massive simulations. The ABM of the arterial wall consists of diseased intima with atherosclerosis, media and adventitia. Figure \ref{fig:vessel_wall} illustrates the three-layered agent-based vessel wall with a placed stent.
For each layer of the ABM of arterial tissue, isotropic tissue with an almost constant density is assumed, which is generated by placing cells randomly, maintaining a minimum distance between neighbouring pairs using a three-dimensional Poisson disc sampling by Bridson's algorithm \cite{Bridson2007}. Interactions occur between adjacent cells, and neo-Hookean and polynomial attractive forces are selected for repulsion and adhesion, respectively \cite{Zun2019}.

It should be noted that an actuall arterial tissue is highly anisotropic. However, we choose to focus on an isotropic formulation as our primary goal is to apply the tissue model to stenting modelling and vessels reconstructed from, e.g. optical coherence tomography or intravascular ultrasound images \cite{Ono2020}. Thus, most of the stress in our use scenarios will likely be in the circumferential direction. Also, finding the fibre directions in an atherosclerotic plaque is technically challenging and not something routinely done - e.g. Akyildiz et al. \cite{Akyildiz2017} get a detailed image of fibres by studying a plaque \emph{ex vivo}. Since this type of data is unexpected for most of the stented vessels, we opt for the model calibration based on the circumferential behaviour of arterial tissue.

\begin{figure}[tb!]
  \centering
  \includegraphics[width=0.55\linewidth]{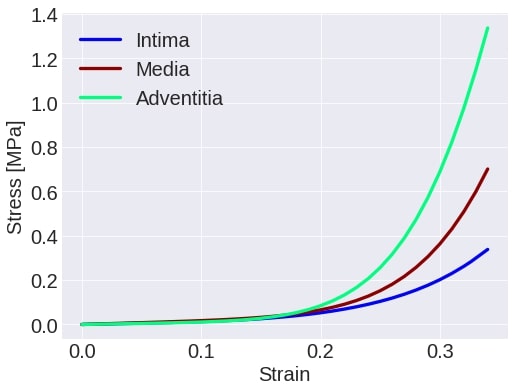}
  \caption{Stress-strain curves obtained by the material model with the $6^{th}$ order reduced polynomial form of the strain energy function for three layers of arterial tissue.
  }
  \label{fig:true_strain_stress}
\end{figure}

\subsection{Mechanical properties of tissue \emph{ex vivo}}
\label{chap:Uniaxial}
One of the most widespread techniques to determine the mechanical properties of arterial tissue is \emph{ex vivo} uniaxial tensile testing. Strips of dissected tissue are subjected to uniaxial tension tests by applying an increasing force to the specimen, and the deformation and ultimate tensile stresses are measured. This provides stress-strain relation for each tested direction of the specimen. The (engineering) strain is defined as the elongation divided by the initial length, and the stress is calculated as the force divided by the cross-sectional area of the specimen \cite{Hayash2000}. In this paper, engineering stress is calculated to match the experimental approach from Holzaphel et al. \cite{Holzapfel2005} and other papers. This way, the initial cross-sectional area is used in the stress calculations. The stress-strain relations for uniaxial strain tests of tissue stretched in a circumferential direction effectively represent the strain the tissue undergoes during the stenting or pressurisation procedures. The computational tissue model calibrated based on the stress-strain behaviour achieving a mechanical response representative of experimental data, can mimic the realistic macroscopic behaviour of arterial tissue.

\subsection{Material model of arterial wall}
\label{subchap:material_model}

The constitutive model for hyperelastic, incompressible rubber-like material is used to produce the ground-truth stress-strain relation of uniaxial tension tests of the arterial tissue for the further calibration task. The generalised polynomial form of the strain energy function is used, which was introduced in the seminal work of Rivlin and Saunders \cite{ RivlinSaunders1951}:
\begin{equation}
    W = \sum_{i=0,j=0}^{\infty} C_{ij}({I_1}-3)^i({I_2}-3)^j 
    \label{eq:GenW}
\end{equation}
where $W$ is a strain energy function and describes the material response while 
$C_{ij}$ is a material parameter that explains the material's shear behaviour.
\noindent ${I}_1 = {\lambda}_1^2 + {\lambda}_2^2 +{\lambda}_3^2  $
and ${I}_2 = \frac{1}{{\lambda}_1^2} + \frac{1}{{\lambda}_2^2} +\frac{1}{{\lambda}_3^2}  $
are two strain invariants, and $\lambda_i$, $i=1,...,3$ is a principal extension ratio, which is in the following relation with the principal strain value $\vartheta$:  $\lambda_i = 1+\vartheta_i$. 

The $6^{th}$ order reduced polynomial strain energy function was employed, which is derived by setting $j=0$ in Equation \ref{eq:GenW}, resulting in the reduced strain energy function to the first strain invariant. The rationale behind reducing the general polynomial function is reviewed by Lapeer et al. \cite{Lapeer2011}. The reduced form of the general polynomial strain energy function has a form \cite{Hartmann2003}:

\begin{equation}
    W = \sum_{i=1}^{N=6} C_{i0}({I_1}-3)^i 
    \label{eq:redU}
\end{equation}
where $N=6$ is the order of the reduced polynomial energy function.  

Since we aim to obtain the stress-strain relation for the uniaxial strain deformation that acts in a single direction, the following associations hold for the principal stretches and stretch in the loading direction - $\lambda$ \cite{RivlinSaunders1951}:

\begin{equation}
    \lambda_1=\lambda  \;\; \text{and} \;\; \lambda_2=\lambda_3=\lambda^{-\frac{1}{2}}, \;\; \text{where} \;\;  \lambda = 1+\vartheta
\end{equation}

The nominal stress for the uniaxial tensile (UT) deformation is obtained by the formula \cite{Treloar1949}:

\begin{equation}
    S_{UT} = \frac{\delta{W}}{\delta{\lambda}}
\end{equation}

As a result, the nominal stress-strain relationship based on the $6^{th}$ order  reduced polynomial strain energy function can be derived :

\begin{equation}
    S_{UT} = 2(\lambda - \lambda^{-2})\sum_{i=1}^{N = 6} i C_{i0}({I_1}-3)^{i-1}
\end{equation}

\begin{table}[tb!]
\centering
\resizebox{.75\columnwidth}{!}{%
\begin{tabular}{c|c|c|c|c|c|c|}
\cline{2-7}
 & \multicolumn{6}{c|}{\textbf{Model parameters}} \\ \cline{2-7} 
\textbf{}                                 & $C_{10}$ & $C_{20}$ & $C_{30}$ & $C_{40}$ & $C_{50}$ & $C_{60}$ \\ \hline
\multicolumn{1}{|c|}{\textbf{Intima}}     & 0.019    & 0.022    & 1.177      & -3.896    & 11.331     & -10.659     \\ \hline
\multicolumn{1}{|c|}{\textbf{Media}}      & 0.0286    & 0.026    & 0.747     & -0.202    & 12.293      & -15.618     \\ \hline
\multicolumn{1}{|c|}{\textbf{Adventitia}} & 0.016    & 0.019    & 1.104      & 3.105      & 21.563      & -41.007     \\ \hline
\end{tabular}
}
\caption{Material model parameters according to \cite{Zun1}, describing three layers of the arterial wall: intima, media and adventitia.}
\label{tab:model_parameters}
\end{table}

Figure \ref{fig:true_strain_stress} shows stress-strain curves obtained by the material model for three layers of arterial tissue.
The choices of the six-order reduced polynomial strain energy function and the material parameters $C_{i0}$, demonstrated in Table~\ref{tab:model_parameters}, are according to the FE arterial model by Zun et al. \cite{Zun1}, which achieved a macroscopic behaviour of an isotropic macroscale model approximating the range of variability of experimental data from a human coronary artery by Holzaphel et al.\cite{Holzapfel2005}. Thus, the stress response of the material model, later used for the model calibration in the IUQ process, is an analytical isotropic approximation of the experimental \emph{in vivo} data.

\subsection{Mechanical properties in agent-based arterial tissue model}
Attractive forces incorporate multiple methods of cell-cell interaction into the model, such as extracellular fibres or cell adhesion. $6^{th}$ order polynomial pairwise interaction force is used as an agent-agent attraction bond force \cite{Zun1}:
\begin{equation}
\centering
\begin{aligned}
F_{bond} = (R_1+R_2)^2 \sum_{i=1}^{6}\theta_i\varrho^i,\;\;\; \text{when}\;\;\;  \varrho > 0 \\ 
\end{aligned}
\end{equation}
 $R_1$ and $R_2$ are the radii of two interacting cells,  $\varrho$ is a bond strain and $\theta_i$, for $i = 1,...,6$ are coefficients of the attraction bond force. The polynomial coefficients are model-specific and do not have any prior meaning or values. The aim is to find those coefficients for the attraction bond force to achieve the macroscopic behaviour of arterial tissue.
 The bond strain  $\varrho$ is quantified by the formula: 
\begin{equation}
    \varrho = \frac{d-R_1-R_2}{R_1+R_2}
\end{equation}
Where $d$ is the distance between the centres of two interacting cells. The choice of the $6^{th}$ order follows the order of the analytical reduced polynomial strain energy function. The corresponding force coefficients are constrained to be non-negative to achieve a minimum of potential energy for each bond at $\varrho=0.$ For different layers of arterial tissue, separate sets of force coefficients and, thus, distinct bond forces are used.
Uniaxial strain tests in the agent-based arterial tissue model are performed separately on intima, media and adventitia in the following manner: a single side of the tissue is fixed, while the rest is stretched up to a particular strain value, and the opposite side is set as well. The system of agents is then solved to find the final shape of the tissue with these boundary conditions. Afterwards, the forces applied to the fixed cells are summed up and divided by the cross-section area to measure the stress for the strain value. Figure \ref{fig:before_linear_force} presents a generated tissue sample of a single layer in an unstrained state, while Figure \ref{fig:after_linear_force} shows the tissue that is subjected to the uniaxial strain, with force coefficients $\theta_1=2.14$ and $\theta_2,\cdots,\theta_6 = 0$. The values are selected to obtain a simple linear fit to the high-strain part of the media data shown in Figure \ref{fig:true_strain_stress}.

The microscopic behaviour of the arterial tissue model strongly depends on the tissue structure, such as the radii of cells, their arrangement, and density of the tissue. 
A detailed investigation of the effect of the density of the tissue on its macroscopic behaviour is presented in the supplementary material - appendix A.

\subsection{Surrogate model}
Here, a Gaussian process regression surrogate model is introduced to represent the latent function $f_{_{GP}}$ that maps input strain values and force coefficients to output stress values. Assume that the stress value $s \in \mathbb{R}$ can be considered as a function of a strain value and polynomial coefficients of attractive force $\boldsymbol{x} = (\theta_1,...\theta_6,\varrho) \in \mathbb{R}^{7}$, for: 
\begin{equation}
     s = f_{_{GP}}(\boldsymbol{x}) + \epsilon, \hspace{0.2cm} \text{where} \hspace{0.1cm} \epsilon \sim \mathcal{N}(0, \sigma_{\epsilon}^2).
\end{equation}
$\epsilon$ is the Gaussian noise term, describing stochasticity in observations with a variance of $\sigma_{\epsilon}^2$.
Given a set of evaluated inputs and outputs of size $N_t$, $\{(\boldsymbol{X},\boldsymbol{s})\} = \{(\boldsymbol{x_i},s_i)\}^{N_t}_{i=1}$, the covariance of $\boldsymbol{s}$ can be written as: 
\begin{equation}
    cov(\boldsymbol{s}) = K(\boldsymbol{X},\boldsymbol{X})+\sigma_{\epsilon}^2 I,
\end{equation}
where  $K(\boldsymbol{X},\boldsymbol{X})$ is a covariance matrix describing correlations between function values at different input points. A Gaussian process is a collection of random variables, any finite collection of which follows a joint multivariate normal distribution \cite{Rasmussen2004}:
\begin{equation}
  f_{_{GP}} \sim \mathcal{GP}(\mu, K), 
\end{equation}
Where $\mu$ and $K$ denote the mean and the covariance functions, respectively, the mean function is generally set to zero to avoid expensive computations. The covariance function, also known as a kernel, contains hyperparameters such as length-scale, signal variance and noise variance. They regulate a priori correlation between arguments. Based on the kernel selection procedure presented in the supplementary material - appendix B - the Matern kernel is selected as a covariance function \cite{Wilkie2021}.

The hyperparameters are fine-tuned by optimising the log marginal likelihood function, which has the following form \cite{Rasmussen2004}:
\begin{equation}
    \log p(\boldsymbol{s}|\boldsymbol{X}) = - \frac{1}{2} \boldsymbol{s}^T (K + \sigma^2_{\epsilon} I)^{-1}\boldsymbol{s} - \frac{1}{2} \log |K+\sigma^2_{\epsilon} I| - \frac{N_{t}}{2}\log 2\pi.
\end{equation}
Subsequently, the evaluations of new unobserved data points are predicted from the resulting posterior distribution. Initially, the prior hyperparameters of the GP model kernel are set to 1 and are optimised using the limited-memory Broyden–Fletcher-Goldfarb–Shanno (L-BFGS) algorithm \cite{Tong2021}. GP regression model is implemented via Gaussian Process framework GPy \cite{GPy}.

\subsubsection{Performance evaluations}
To evaluate the predictive capability of a surrogate model, we assess the prediction accuracy of the surrogate via the root mean square error (RMSE), standardised root mean squared error (SRMSE) and the coefficient of determination (also referred to as $R^2$ ). $R^2$ is a ratio of the output variance explained by the surrogate model \cite{Chicco2021}, and its value ranges from 0 (null predictivity) to 1 (perfect predictivity).

\subsubsection{Experimental design}
Separate layers of arterial tissue have different mechanical properties \cite{Holzapfel2005} and are represented by three distinct interaction forces and are calibrated individually. Thus, separate sets of coefficients are found for intima, media and adventitia. Three surrogate models for each layer of arterial tissue were developed for IUQ.

\begin{table}[tb!]
\centering
\resizebox{0.75\columnwidth}{!}{%
\begin{tabular}{c|ccc|ccc|ccc|cc|}
\cline{2-12}
\textbf{}                                 & \multicolumn{3}{c|}{\textbf{Intima}}                                              & \multicolumn{3}{c|}{\textbf{Media}}                                               & \multicolumn{3}{c|}{\textbf{Adventitia}}                                          & \multicolumn{2}{c|}{\textbf{Joint}}            \\ \cline{2-12} 
                                          & \multicolumn{1}{c|}{\textbf{PS}} & \multicolumn{1}{c|}{\textbf{LB}} & \textbf{UB} & \multicolumn{1}{c|}{\textbf{PS}} & \multicolumn{1}{c|}{\textbf{LB}} & \textbf{UB} & \multicolumn{1}{c|}{\textbf{PS}} & \multicolumn{1}{c|}{\textbf{LB}} & \textbf{UB} & \multicolumn{1}{c|}{\textbf{LB}} & \textbf{UB} \\ \hline
\multicolumn{1}{|c|}{\boldsymbol{$\theta_1$}} & \multicolumn{1}{c|}{0.2}         & \multicolumn{1}{c|}{0}           & 2           & \multicolumn{1}{c|}{0.2}         & \multicolumn{1}{c|}{0}           & 2           & \multicolumn{1}{c|}{0.2}         & \multicolumn{1}{c|}{0}           & 2           & \multicolumn{1}{c|}{0}           & 2           \\ \hline
\multicolumn{1}{|c|}{\boldsymbol{$\theta_2$}} & \multicolumn{1}{c|}{1}           & \multicolumn{1}{c|}{0}           & 2           & \multicolumn{1}{c|}{1}           & \multicolumn{1}{c|}{0}           & 2           & \multicolumn{1}{c|}{1}           & \multicolumn{1}{c|}{0}           & 2           & \multicolumn{1}{c|}{0}           & 2           \\ \hline
\multicolumn{1}{|c|}{\boldsymbol{$\theta_3$}} & \multicolumn{1}{c|}{2}           & \multicolumn{1}{c|}{0}           & 4           & \multicolumn{1}{c|}{2}           & \multicolumn{1}{c|}{0}           & 4           & \multicolumn{1}{c|}{2}           & \multicolumn{1}{c|}{0}           & 4           & \multicolumn{1}{c|}{0}           & 4           \\ \hline
\multicolumn{1}{|c|}{\boldsymbol{$\theta_4$}} & \multicolumn{1}{c|}{3}           & \multicolumn{1}{c|}{0}           & 4           & \multicolumn{1}{c|}{3}           & \multicolumn{1}{c|}{0}           & 6           & \multicolumn{1}{c|}{3}           & \multicolumn{1}{c|}{0}           & 6           & \multicolumn{1}{c|}{0}           & 6           \\ \hline
\multicolumn{1}{|c|}{\boldsymbol{$\theta_5$}} & \multicolumn{1}{c|}{4}           & \multicolumn{1}{c|}{0}           & 8           & \multicolumn{1}{c|}{4}           & \multicolumn{1}{c|}{0}           & 12          & \multicolumn{1}{c|}{4}           & \multicolumn{1}{c|}{0}           & 12          & \multicolumn{1}{c|}{0}           & 12          \\ \hline
\multicolumn{1}{|c|}{\boldsymbol{$\theta_6$}} & \multicolumn{1}{c|}{500}         & \multicolumn{1}{c|}{300}         & 700         & \multicolumn{1}{c|}{1600}        & \multicolumn{1}{c|}{1200}        & 1800        & \multicolumn{1}{c|}{3200}        & \multicolumn{1}{c|}{2800}        & 3500        & \multicolumn{1}{c|}{300}         & 3500        \\ \hline
\end{tabular}
}
\caption{The preliminary sets (PS) of the attractive force coefficients and their lower (LB) and upper (UB) bounds of the sampling ranges for three layers (intima, media and adventitia) of the arterial tissue model and "Joint" bounds covering all three ranges.}
\label{tab:coeff_ranges_restricted}
\end{table}

ABM stress responses for uniaxial strain tests on an extensive collection of sets of coefficients were obtained to gather sufficient data for building surrogate models. A uniaxial strain test is performed by stretching the tissue to the strain value varied from 0 to 0.35 with an increment of 0.005, giving in a total number of observations $N_{tot} = 71$ of stress-strain values.
Table \ref{tab:coeff_ranges_restricted} shows preliminary sets of force coefficients as an initial guess for intima, media and adventitia layers, such that when given to the ABM, they produce stress-strain curves in a rough agreement with the benchmark data in terms of magnitude. The training ranges of coefficients for three layers and "Joint" bounds covering all three ranges are demonstrated in Table \ref{tab:coeff_ranges_restricted}. All the upper and lower bounds for each layer are selected around the preliminary guess, which is justified by the sensitivity analysis of each coefficient, demonstrated in the supplementary material - appendix C. 

Sets of coefficients $\boldsymbol{\theta} = (\theta_1,...\theta_6)$ were individually sampled from the designed ranges using Latin Hypercube Sampling (LHS) for intima, media and adventitia, leading to three separate input spaces. LHS, used to achieve good parameter space coverage and avoid clustering,  was implemented via an open-source Python package SMT \cite{SMT2019}.
The validation data contains a large number of input parameter sets for all strain values, resulting in stress outcomes of different magnitudes, which would result in high sensitivity of error measures to the scale of target values. To tackle this issue, RMSE was calculated for each strain value separately and was normalised by the variance of the target stress values \cite{Rasmussen2004}, giving SRMSE values, which were finally summed up to obtain the total SRMSE term for each layer. 

\subsection{Inverse uncertainty quantification}
\subsubsection{Bayesian calibration}
Bayesian calibration relates prior information with uncertainty to posterior information based on the likelihood of simulated outputs from the computational model. In each iteration of the Bayesian calibration, the posterior probability density functions of calibration parameters are updated in a way that is most likely to align with the benchmark data. The posterior distributions of the parameters inferred after a sufficient number of iterations are the most probable calibration of the model, which means that a model calibrated with Bayesian calibration can efficiently produce expected behaviour by determining the best estimate of parameter uncertainties.

We aim to calibrate the set of attractive force parameters $\boldsymbol{\theta}$ based on the evidence data of stress output $D$ during the stress-strain tests.
Bayesian calibration is an application of the Bayesian inference method, in which the probability for a hypothesis is updated as more evidence is provided:
\begin{equation}
    p(\boldsymbol{\theta}|D) = \frac{p(\boldsymbol{\theta},D)}{p(D)} = \frac{p(\boldsymbol{\theta}) p(D|\boldsymbol{\theta})}{p(D)}
    \label{Baye}
\end{equation}
where $p(\boldsymbol{\theta})$ is a prior distribution of the model parameters $\boldsymbol{\theta}$, $p(D|\boldsymbol{\theta})$ is known as \textit{likelihood} or \textit{sampling distribution} and $p(D) = \int_{ \Theta} p(\boldsymbol{\theta})p(D|\boldsymbol{\theta})d\boldsymbol{\theta}$ is integrated over the full parameter space $ \Theta$. However, since  $p(D)$ does not depend on $\boldsymbol{\theta}$ and for fixed $D$ it can be considered as a constant, equation \ref{Baye} can be reformulated as \emph{unnormalized posterior density} \cite{Gelman2013}: 
\begin{equation}
    \centering
    p(\boldsymbol{\theta}|D) \propto p(D|\boldsymbol{\theta}) \; p(\boldsymbol{\theta})
    \label{eq:inference}
\end{equation}

\subsubsection{IUQ model formulation}
The iterative Bayesian calibration process is conducted based on the model updating equation - "true value = simulated value + uncertainty", which links observation values to the model response and uncertainty and is used to tune unknown model calibration parameters and the discrepancy term. 
In this work, a model updating equation without a model discrepancy function was selected since combining calibration of the model parameters and the discrepancy function faces the non-identifiability problem due to the challenging task of jointly fitting the model and model discrepancy. The term refers to the state when the model choice and the selected calibration targets are insufficient to obtain the unique values of the calibration parameters \cite{Escudero2018}. Besides, in most cases, it is difficult to distinguish the effects of the discrepancy function and calibration parameters on response predictions when both are incorporated in the calibration process \cite{Arendt2012}.
We introduce a zero-mean Gaussian bias term that accounts for aleatory and epistemic uncertainty sources, and its variance is calibrated throughout the IUQ process \cite{Wu2018InverseTheory, deVries2020}.

Inverse uncertainty quantification is formulated using the terminology of Kennedy and O'Hagan \cite{Kennedy2001}. The model has variable and calibration inputs. The variable inputs have known values for each observation used in IUQ, while the calibration inputs are the unknown parameters we aim to identify. The variable inputs $\boldsymbol{\varrho}=(\varrho_{1},...,\varrho_{N_{obs}})$ are $N_{{obs}}$ observed strain values, obtained by stretching the tissue so that the corresponding strain value varies from $0$ to $0.35$ by an increment of $0.025$. $\boldsymbol{\theta}_i=(\theta_{1, i},...,\theta_{6, i})$, $i=1,..., N_{s}$  is a vector of unknown attractive force coefficients we aim to calibrate, where $N_{s}$ is a size of sampled coefficient space, consisting of sets of different combinations of coefficient values. 
 
The benchmark data used for calibration is the analytical stress response of uniaxial stress-strain tests, obtained by the material model and is denoted by $\textbf{z} = (z_1,..,z_{N_{obs}})$. The IUQ model stress responses for sets of the input parameters 
$(\boldsymbol{\theta}_i, \boldsymbol{\varrho})$, for $i=1,..., N_{s}$ 
are produced as predictions of the Gaussian process surrogate model,  denoted by $\textbf{s} = (s_1 ,..,s_{{N_p}})$, where $N_p$ is a total number of predictions. 
The set of stress data for the IUQ process is then denoted by $D = (\textbf{z},\textbf{s})$. As a result, the following relation is inferred between the analytical values of stress and the ABM response:
\begin{equation}
    \boldsymbol{z} = f_{ABM}(\boldsymbol{\theta^*},\boldsymbol{\varrho}) + \varepsilon
\end{equation}
\noindent where $f_{_{ABM}}$ - describes the original arterial tissue ABM we are calibrating, $\varepsilon$ is a total uncertainty of the IUQ model, while $\boldsymbol{\theta^*}$ is the reference value of calibrated parameters and the quantity of interest, such that given to the ABM, stress response for the uniaxial strain tests aligns with the analytical solution.

\subsubsection{Modelling prediction uncertainty}
The $\varepsilon$ term represents a total \emph{prediction uncertainty}, which can be categorised into the four main groups:  \emph{observation error},  \emph{residual variability}, \emph{code uncertainty}  and \emph{model inadequacy} \cite{Kennedy2001}. The observation error is the uncertainty caused by the measurement noise when collecting the experimental data. At the same time, the residual variability is the aleatory uncertainty of the model when parameters and conditions are fully specified and fixed. The code uncertainty comes from the fact that ABM response at any given set of inputs used in IUQ can not be obtained due to the high computational costs; thus,  metamodel for predicting stress value in unknown input points is introduced, which raises code or also known as $interpolation \; uncertainty$. The model inadequacy also called $model \;bias$ can be caused by missing physics, inaccurate modelling assumptions, numerical errors or any other causes that can not be assessed in advance in contrast with the error terms mentioned earlier.
Since our calibration data is a simulation output of the material model of arterial tissue and not measurements from \emph{in vivo} or \emph{in vitro} experiments, observation error can be discarded. We only consider residual variability $\varepsilon_r$, the model inadequacy $\varepsilon_m$ and the code uncertainty $\varepsilon_c$.

The total uncertainty is broken into three uncertainty components and is represented as their sum, following De Vries et al. \cite{deVries2020}:
\begin{equation}
    \varepsilon = \varepsilon_r +\varepsilon_c +\varepsilon_m 
\end{equation}
Uncertainty terms are assumed to follow independent Gaussian distributions with zero mean, and corresponding covariance matrices:
\begin{equation}
    \varepsilon_r \sim \mathcal{N}(0, \Sigma_{{\varepsilon_r}}), \;\;
    \varepsilon_c \sim \mathcal{N}(0, \Sigma_{{\varepsilon_c}}), \;\;
    \varepsilon_m \sim \mathcal{N}(0, \Sigma_{{\varepsilon_m}})
\end{equation}
Besides, the error variables are assumed to be independently and identically distributed. Thus, they are uncorrelated and their covariance matrices have diagonal structures, with the diagonal elements representing the variances and the off-diagonal elements equal to zero. Due to the independent Gaussian distribution of uncertainties, the total prediction error term $\varepsilon$ has the Gaussian distribution and its covariance matrix has the following form:
\begin{equation}
\begin{aligned}
     \varepsilon \sim \mathcal{N}(0, \Sigma_{{\varepsilon}}), \;\;\;\;\;\;\;\;  \\
\centering
  \Sigma_{\varepsilon} =\Sigma_{\varepsilon_r}+\Sigma_{\varepsilon_c} + \Sigma_{\varepsilon_m}
\end{aligned}
\end{equation}
As mentioned above,  $\varepsilon_r$ and  $\varepsilon_c$ are known a priori. 
Since stochasticity of the ABM or GP prediction uncertainty is not prevalent, for simplicity, homoscedastic error terms with equal variances are assumed: 
\begin{equation}
    \Sigma_{\varepsilon_r} = \sigma_r^2 \mathcal{I}, \;\;\;\; \Sigma_{\varepsilon_c} = \sigma_c^2 \mathcal{I}, \;\;\;\; \Sigma_{\varepsilon_m} = \sigma_m^2 \mathcal{I} 
\end{equation}
where $\mathcal{I}$ is an identity matrix,  $\sigma_r^2$ is a mean variance of 100 simulations of ABM stress response for uniaxial strain tests per strain values, while $\sigma_c^2$ is averaged GP's prediction variance for the validation data. The variance parameter $\sigma_m^2$, used for modelling uncertainties in $\varepsilon_m$, is estimated along the IUQ process. As a result, the full set of uncertain calibration parameters is defined as $\Phi =(\boldsymbol{\theta}, \sigma_m^2)$, where  $\boldsymbol{\theta}$ and $\sigma_m^2$ are assumed to be independently distributed. 
The likelihood is modelled as a joint Gaussian distribution \cite{Rasmussen2004}. Its mean is calibration data given the GP model response for the prior uncertain parameters, and the covariance matrix $ \Sigma_{\varepsilon} $ is from the total uncertainty $ \varepsilon $:
\begin{equation}
    \mathcal{L}( D|\Phi) = \mathcal{N}(\boldsymbol z|\; f_{gp}(\boldsymbol\theta,\boldsymbol\varrho), \; \boldsymbol{\Sigma_{{\varepsilon}}} )
\end{equation}
\begin{equation*}
  \mathcal{L}( D|\Phi) = \;\frac{|\boldsymbol{\Sigma_{\varepsilon}}|^{-\frac{1}{2}} }{2\pi^{\frac{N_s}{2}}} \;\text{exp}\; \Big(-\frac{1}{2}(\boldsymbol z-f_{gp}(\boldsymbol\theta, \boldsymbol \varrho))^T \boldsymbol{\Sigma^{-1}_{\varepsilon}}(\boldsymbol z-f_{gp}(\boldsymbol{\theta},\boldsymbol{\varrho}))\Big)
\end{equation*}
Since optimal parameter ranges were not known in advance,  uniform distribution was used to obtain the prior marginal distributions. The lower and upper bounds of uniform distribution were according to Table \ref{tab:coeff_ranges_restricted}.

\subsubsection{Calibration process}
After defining prior distributions for the calibration parameters and the likelihood function, a MCMC sampling with a Metropolis-Hastings algorithm was used to sample from the obtained posterior distributions \cite{Hastings1970}. 
An initial state for the sampler in the parameter space by the maximum a posteriori (MAP) method was found, which is a numerical optimisation method to find a point estimate of the mode of the distribution \cite{Bassett2019}. The calibration process was then performed iteratively to update our beliefs about the calibration parameters. The obtained posterior distributions were used as the following inference's prior distributions, and new posteriors were produced. This process of the calibration was then iterated until the posterior distributions converged. For reusing posteriors, first, Gaussian kernel density estimation (KDE) \cite{Wang2020} was used to estimate the PDF of a random variable in a non-parametric way. To ensure convergence to true parameter values, independent data is needed in each iteration. Thus, for sampling from obtained PDFs, a linear interpolation of PDF was evaluated on evenly distributed points on the extended domain of the posterior samples. The Bayesian calibration process is performed using a probabilistic programming package for Python - PyMC3 \cite{PyMC3}. 

\section{Results}\label{cha:results}

\subsection{Performance of GP}
Initially, a single Gaussian process model was trained on the parameter space constructed with the $N_{obs} = 15$ strain values and $1000$ sets of force coefficients sampled from the "Joint" ranges given in Table \ref{tab:coeff_ranges_restricted}. The ratio between a training and validation set was 70\% to 30\%. Gaussian process predictions were compared to the ABM stress responses on the validation data for the complete set of strain values consisting of $N_{tot} = 71$ observations; thus, interpolation performance in unknown strain values was also monitored. The quality of the surrogate predictor was assessed via the total SRMSE and the coefficient of determination $R^2$ on the validation data. As demonstrated in Table \ref{tab:gp_pred_measures}, values of total SRMSEs are low, and $R^2$ coefficients are close to 1.0, which suggests satisfactory predictive capabilities of the surrogate. The investigation of the mean difference between GP prediction with a 95\% confidence interval and ABM stress response is presented in the supplementary material - Appendix D.

\begin{table}[tb]
\centering
\resizebox{0.55\columnwidth}{!}{
\begin{tabular}{c|c|c|c|}
\cline{2-4}
& \textbf{Intima} & \textbf{Media} & \textbf{Adventitia} \\ \hline
\multicolumn{1}{|c|}{\textbf{ Total SRMSE}} & 0.0007              & 0.004    &    0.01              \\ \hline
\multicolumn{1}{|c|}{\boldsymbol{$R^2$}} & 0.99998               & 0.99996          & 0.99995                   \\ \hline
\end{tabular}
}
\caption{Total SRMSE and predictivity coefficient $R^2$ of GP surrogate model for three layers of the arterial tissue model. }
\label{tab:gp_pred_measures}
\end{table}

\begin{figure}[tbh]
 \centering
\includegraphics[width=0.55\linewidth]{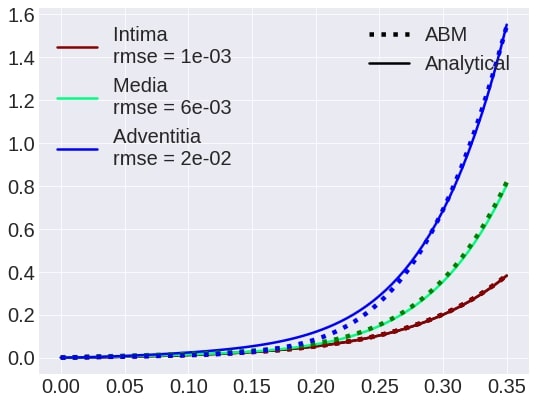}
\caption{GP prediction with 95\% CI versus ABM stress response with the preliminary set of force coefficients for intima, media and adventitia. 
} 
\label{fig:GP_VS_ABM_Restricted_Separates}
\end{figure}

For an intelligible illustration of the predictive capabilities of the surrogate, stress-strain curves produced by the ABM and GP model predictions given the set of preliminary coefficients and the full set of strain values were compared to each other. Figure \ref{fig:GP_VS_ABM_Restricted_Separates}  illustrates this comparison for intima, media and adventitia layers. In all three cases, the prediction of the surrogate model is in line with ABM stress behaviour, producing low RMSEs and narrow confidence intervals of predictions. 

\subsection{Inverse uncertainty quantification}
\begin{figure}[tb!]
\centering

\begin{subfigure}{.32\textwidth}
  \centering
  \includegraphics[width=\linewidth]{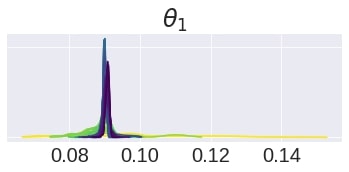}  
  \label{fig:intima-posteriors1}
\end{subfigure}
\vspace{-1\baselineskip}
\begin{subfigure}{.32\textwidth}
  \centering
  \includegraphics[width=\linewidth]{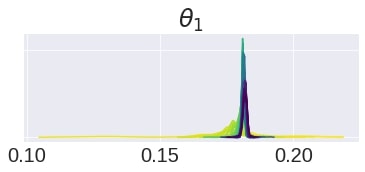}  
  \label{fig:media-posteriors1}
\end{subfigure}
\begin{subfigure}{.32\textwidth}
  \centering
  \includegraphics[width=\linewidth]{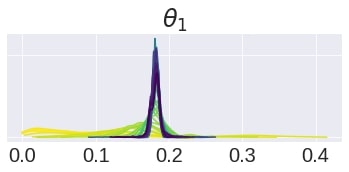}  
  \label{fig:adv-posteriors1}
\end{subfigure}
\vspace{-1\baselineskip}
\begin{subfigure}{.32\textwidth}
  \centering
  \includegraphics[width=\linewidth]{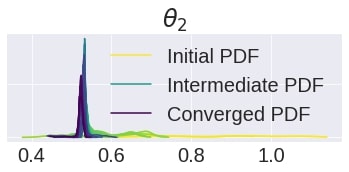}  
  \label{fig:intima-posteriors2}
\end{subfigure}
\begin{subfigure}{.32\textwidth}
  \centering
  \includegraphics[width=\linewidth]{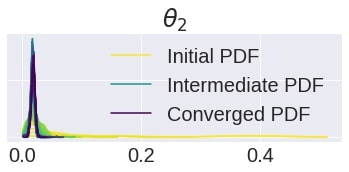}  
  \label{fig:media-posteriors2}
\end{subfigure}
\begin{subfigure}{.32\textwidth}
  \centering
  \includegraphics[width=\linewidth]{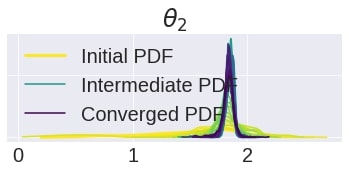}  
  \label{fig:adv-posteriors2}
\end{subfigure}
\vspace{-1\baselineskip}
\begin{subfigure}{.32\textwidth}
  \centering
  \includegraphics[width=\linewidth]{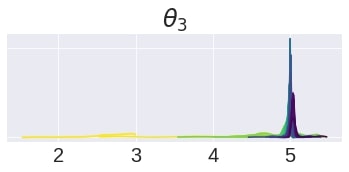}  
  \label{fig:intima-posteriors3}

\end{subfigure}
\begin{subfigure}{.32\textwidth}
  \centering
  \includegraphics[width=\linewidth]{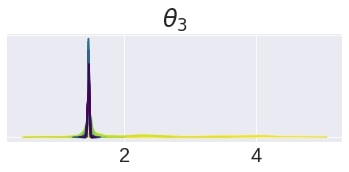}  
  \label{fig:media-posteriors3}
\end{subfigure}
\begin{subfigure}{.32\textwidth}
  \centering
  \includegraphics[width=\linewidth]{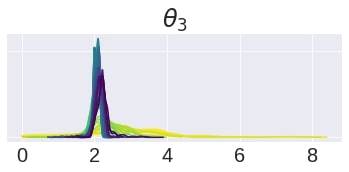}  
  \label{fig:adv-posteriors3}
\end{subfigure}
\vspace{-1\baselineskip}
\begin{subfigure}{.32\textwidth}
  \centering
  \includegraphics[width=\linewidth]{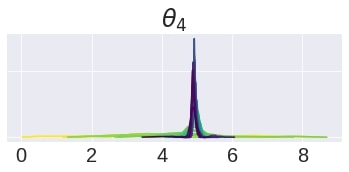}  
  \label{fig:intima-posteriors4}
\end{subfigure}
\begin{subfigure}{.32\textwidth}
  \centering
  \includegraphics[width=\linewidth]{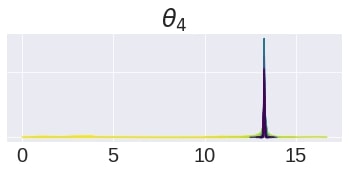}  
  \label{fig:media-posteriors4}
\end{subfigure}
\begin{subfigure}{.32\textwidth}
  \centering
  \includegraphics[width=\linewidth]{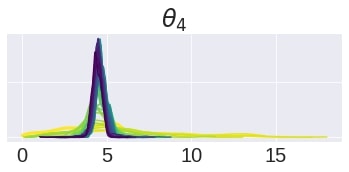}  
  \label{fig:adv-posteriors4}
\end{subfigure}
\vspace{-1\baselineskip}
\begin{subfigure}{.32\textwidth}
  \centering
  \includegraphics[width=\linewidth]{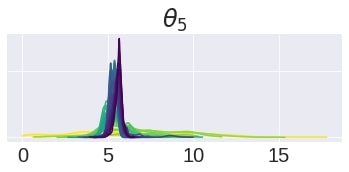}  
  \label{fig:intima-posteriors5}
\end{subfigure}
\begin{subfigure}{.32\textwidth}
  \centering
  \includegraphics[width=\linewidth]{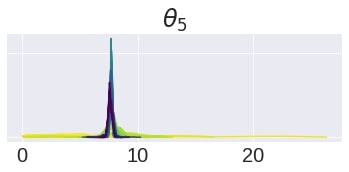}  
  \label{fig:media-posteriors5}
 \end{subfigure}
\begin{subfigure}{.32\textwidth}
  \centering
  \includegraphics[width=\linewidth]{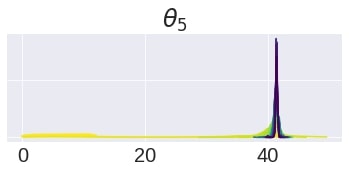}  
  \label{fig:adv-posteriors5}
\end{subfigure}
\vspace{-1\baselineskip}
\begin{subfigure}{.32\textwidth}
  \centering
  \includegraphics[width=\linewidth]{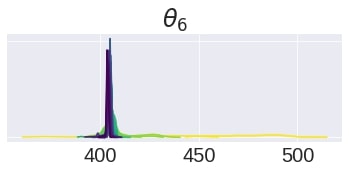}  
  \label{fig:intima-posteriors6}
\end{subfigure}
\begin{subfigure}{.32\textwidth}
  \centering
  \includegraphics[width=\linewidth]{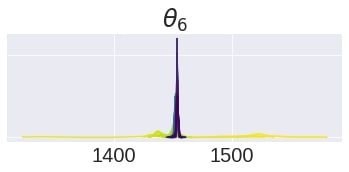}  
  \label{fig:media-posteriors6}
\end{subfigure}
\begin{subfigure}{.32\textwidth}
  \centering
  \includegraphics[width=\linewidth]{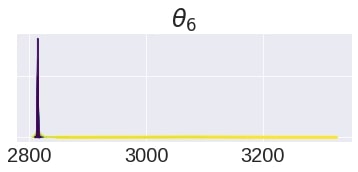}  
  \label{fig:adv-posteriors6}
\end{subfigure}
\begin{subfigure}{.32\textwidth}
  \centering
  \includegraphics[width=\linewidth]{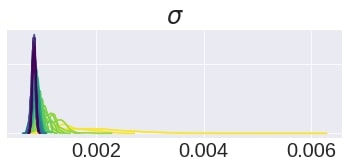}  
  \caption{Intima}
  \label{fig:intima-posteriors7}
\end{subfigure}
\begin{subfigure}{.32\textwidth}
  \centering
  \includegraphics[width=\linewidth]{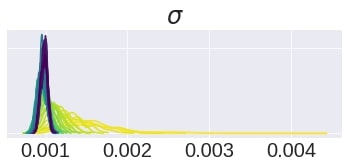}  
  \caption{Media}
  \label{fig:media-posteriors7}
\end{subfigure}
\begin{subfigure}{.32\textwidth}
  \centering
  \includegraphics[width=\linewidth]{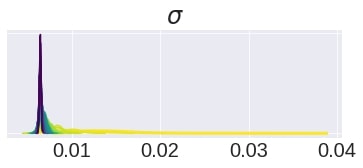}  
  \caption{Adventitia}
  \label{fig:adv-posteriors7}
  \end{subfigure}
\caption{1D marginal posterior distributions of the calibration parameters for the intima, media and adventitia layers of arterial tissue after 80 iterations of Bayesian calibration.}
\label{fig:intima-posteriors}
\end{figure}

Inverse uncertainty quantification, using Bayesian calibration of the polynomial force coefficients and the model uncertainty term, was conducted for three layers of the arterial tissue. Figure \ref{fig:intima-posteriors} illustrate PDFs of model parameters during the Bayesian calibration process. PDFs from earlier calibration steps are nearly flat, close to a prior uniform distribution, referring to flat marginal likelihoods of the model parameters, which means that there is a big number of maximum likelihood estimates of parameters. In later iterations, we see a reduction in uncertainty and posterior PDFs that turn into narrow probability distributions, far from the prior, signifying model identifiability. Finally,  a convergence of marginal posterior distributions of the parameters to the targeted distributions becomes apparent for three layers after 80 iterations of Bayesian calibration. The calibration process on the plots is presented via the sequential colour scheme, where initial PDFs are shown by light yellow colours, which are monotonically getting darker throughout the calibration process and finally converge to targeted posterior distributions presented with the dark blue colour.

\begin{table}[tb!]
\centering
\resizebox{0.75\columnwidth}{!}{%
\begin{tabular}{c|c|c|c|c|c|c|}
\cline{2-7}
\textbf{}                                & \multicolumn{2}{c|}{\textbf{Intima}} & \multicolumn{2}{c|}{\textbf{Media}} & \multicolumn{2}{c|}{\textbf{Adventitia}} \\ \hline
\multicolumn{1}{|c|}{\textbf{Parameter}} & \textbf{Mean}      & \textbf{sd}     & \textbf{Mean}     & \textbf{sd}     & \textbf{Mean}        & \textbf{sd}       \\ \hline
\multicolumn{1}{|c|}{$\sigma_{\epsilon_m}$}  &   0.001 & 0.000 &    0.001 & 0.000 &    0.006 & 0.000 \\ \hline
\multicolumn{1}{|c|}{\textbf{$\theta_1$}}    &   0.091 & 0.001 &    0.182 & 0.001 &    0.181 & 0.006 \\ \hline
\multicolumn{1}{|c|}{\textbf{$\theta_2$}}    &   0.526 & 0.011 &    0.019 & 0.003 &    1.838 & 0.044 \\ \hline
\multicolumn{1}{|c|}{\textbf{$\theta_3$}}    &   5.008 & 0.077 &    1.454 & 0.040 &    2.216 & 0.139 \\ \hline
\multicolumn{1}{|c|}{\textbf{$\theta_4$}}    &   4.976 & 0.244 &   13.277 & 0.087 &    4.399 & 0.297 \\ \hline
\multicolumn{1}{|c|}{\textbf{$\theta_5$}}    &   5.672 & 0.259 &    7.607 & 0.209 &   41.461 & 0.136 \\ \hline
\multicolumn{1}{|c|}{\textbf{$\theta_6$}}    & 403.626 & 0.558 & 1453.614 & 0.868 & 2814.243 & 0.493 \\ \hline
\end{tabular}
}
\caption{Statistics of converged posterior distributions for three layers of arterial tissue, namely, the mean and standard deviation of the resulted probability distributions after 80 iterations of Bayesian calibration.  }
\label{tab:iuq_statistics}
\end{table}

Table \ref{tab:iuq_statistics} shows the statistics of the converged posterior distributions of 6 attractive force coefficients and the variance of the model inadequacy term after 80 iterations for three layers of the tissue. More specifically, the mean and standard deviation of the resulted probability distributions. 
Besides, the potential scale reduction factors (PSRF) were calculated, which can be considered as a convergence diagnostic \cite{Roy2020}. PSRF values were very close to $1$ for each layer and calibration parameter, indicating that associated chains likely converged to the targeted posterior distributions \cite{Jones2022}. Thus, running simulations any longer was considered unnecessary. 

The marginal distribution of $\sigma_{\varepsilon_m}$, the variance of the model uncertainty term $\varepsilon_m$, is centred around the mean of $0.001$, which validates the model and the assumption that the prediction error is normally distributed with a zero mean. For assessing the precision of the IUQ results, mean values of the final marginal posterior distributions of the coefficients were used for collecting the GP predictions and ABM responses per the specified set of force coefficients and the complete set of observed strain values.

\begin{figure}[tb!]
 \centering
\includegraphics[width=0.55\linewidth]{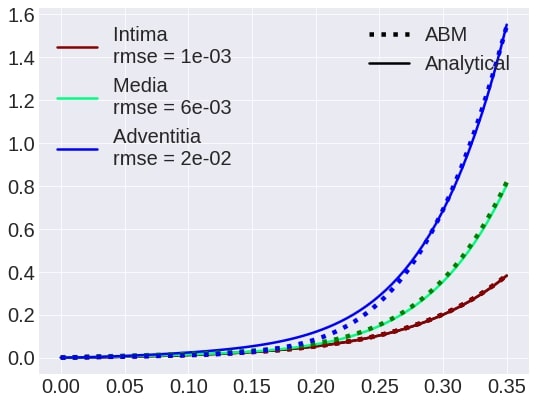}
\caption{ABM stress response for uniaxial strain tests with the calibrated set of force coefficients,  given in Table \ref{tab:iuq_statistics}, versus analytical data for three layers of arterial tissue. }
\label{fig:calibration_comparision}
\end{figure}

Figure \ref{fig:calibration_comparision} illustrates the obtained stress-strain curves of GP and ABM compared to the analytical data of the material model. The ABM of arterial tissue shows sufficient precision in fitting the ground-truth uniaxial strain test data with RMSE value up to $2.1\%$, which verifies that IUQ procedure was successful and the aim of the ABM to replicate realistic mechanical properties was achieved. 

\subsection{Speed up}

The ABM of arterial tissue is implemented using OpenMP, which supports shared-memory, parallel multiprocessing programming. The uniaxial strain tests were run on the 16-core node of the SURFsara Lisa cluster in parallel, utilising all 16 cores for each test. The average simulation time of a single uniaxial strain test ranges between 1.3 to 1.5 minutes wall clock and depends on the scale of the force coefficient values, taking longer time for the higher values and vice versa. One Bayesian calibration iteration of the IUQ for satisfactory outcomes requires at least 4000 draws of the model stress response. Thus, a single calibration iteration for one out of three arterial layers would take 96 hours if running the original model. On the other hand, the GP model takes 0.2 milliseconds on average to produce the stress prediction, irrespective of the magnitude of input coefficient values. 

\section{Discussion} \label{cha:discussion}

The benchmark data every IUQ and calibration problem relies on was obtained analytically by following an isotropic material model of arterial wall \cite{Zun1}, which is a representation of \emph{in vitro} behaviour of arterial tissue, approximating the range of variability of the experimental stress-strain curves \cite{Holzapfel2005}. Flexibility to generate desirable data allowed for efficient calibration and validation processes. GP regression approximated the original model's stress behaviour in response to the uniaxial tissue stretching in a circumferential direction. Separate surrogate models for intima, media and adventitia were obtained, which provides generality and compatibility of the method to future layer-specific arterial tissue modelling modifications, as well as similarly structured models of other biological tissues. Obtaining a precise and computationally efficient representation of the ABM's mechanical behaviour was vital for effective sensitivity analysis and calibration processes.

A Bayesian calibration with a bias term correction was performed as a technique of IUQ to quantify uncertainties of attractive force coefficients and estimate their values based on the analytical data. 
Since our objective was to find the precise values of the polynomial force function coefficients, the challenge of distinguishing between the effects of calibration parameters and other factors could have been limited. Besides, as the coefficients do not have biological value, the limitation of the possibly converging to "pseudo-true" values, lacking physical interpretation, was not much of a concern. 

When attractive forces were formed based on the calibrated coefficients, performing uniaxial strain tests on the ABM showed a stress-strain relation in line with analytical data for all three layers. As a result, uncertainties about the unknown parameters were reduced, and the attractive forces of the ABM of arterial tissue layers were determined, providing a reasonable macroscopic stress-strain relationship for uniaxial strain tests.
The efficient computational model of arterial tissue can ensure the reliable and accurate behaviour of a larger-scale model with a broader scope and diverse applications, of which it is a component. Specifically, the microscale mechanical model can be used to provide mechanical information to a biological model of cells on a microscale. 
The direct application of the obtained arterial tissue model with realistic macroscopic behaviour will simulate the implantation of a stent for the ISR3D model. 
The paper by Zun et al. \cite{Zun2019} shows the application of ISR3D by modelling the In-stent restenosis (ISR) process in porcine coronary arteries and validating the results by \emph{in vivo} data. Furthermore, the arterial tissue model can contribute to advancing vascular medicine and clinical treatment of artery diseases, designing interventional devices such as vascular implants \cite{Perkins2019} and limiting unfeasible, expensive and sometimes unethical \emph{in vivo} or \emph{in vitro} studies. A viable microscale agent-based model of arterial tissue can be adapted to other mechanobiological applications, such as models of valves or vein grafts.

A limitation of the arterial tissue model is that only the isotropic formulation is considered and calibrated to match the circumferential behaviour of the vessel. However, extending the model to include anisotropy is in our future plans. It should also be noted that the approach outlined in this paper can be generalised to other pairwise interaction forces, including anisotropic ones if uniaxial strain tests in the axial and radial directions are also included. 
The methodology presented here can be applied to other microscale models where mechanical properties are important. Furthermore, according to the complexity of the phenomenon and dynamics of interest, more elaborate methodological approaches can be incorporated, such as modelling inadequacy function when the IUQ parameters have physical values \cite{Maupin2020}; using the active learning strategy for the surrogate modelling \cite{Zhang2019}; reducing dimensionality incorporating principal components that account for uncertainty of the high-dimensional dynamic output {\cite{Liu2021_2}}; and/or calibrating the parameters of the surrogate model in the IUQ process \cite{Wu2018IUQ2} if the pre-trained surrogate model does not achieve the sufficient precision.

\section{Conclusions}\label{cha:conclusion}

This paper applied inverse uncertainty quantification to determine interaction forces in the cell-resolved agent-based arterial tissue model from the analytical data of biological tissue's macroscopic behaviour. Considering the computational intensity of the model, the necessity of surrogate modelling was anticipated. Overall, attractive force coefficients were successfully calibrated using the proposed IUQ model, which means that interaction forces between the agent-based arterial tissue model cells were found, providing a reasonable macroscopic stress-strain relationship for uniaxial strain tests. As a result, the model reveals a realistic mechanical behaviour of biological tissue.

The result is a versatile and generalisable approach for modelling and calibrating microscale model of natural phenomena using inverse uncertainty quantification techniques and surrogate modelling based on a macroscale mechanical model.

\section{Contributions}
\noindent Salome Kakhaia: Conceptualisation, Methodology, Software, Writing - original draft. Pavel Zun: Conceptualisation, Methodology, Writing - Review \& Editing. Dongwei Ye: Conceptualisation, Methodology, Writing - Review \& Editing. Valeria Krzhizhanovskaya:
Conceptualisation, Writing - Review \& Editing, Supervision.

\section{Funding}
\noindent This project has received funding from the European Union Horizon 2020 research and innovation programme under grant agreements \#800925 (VECMA project), \#777119 (InSilc project), \#101016503 (In Silico World project). PZ acknowledges funding from the Russian Science Foundation under agreement \#20-71-10108 and from ITMO University under agreement 621291. This work was sponsored by NWO Exacte Wetenschappen (Physical Sciences) for the use of supercomputer facilities, with financial support from the Nederlandse Organisatie voor Wetenschappelijk Onderzoek (Netherlands Organization for Science Research, NWO). 

\section{Acknowledgements}
\noindent We thank L. Antonini for the information and discussions on continuous models of arterial tissue.

\label{References}
\bibliographystyle{IEEEtran}
\bibliography{References_new.bib}


\end{document}